\documentstyle[epsf]{mn}
\newcommand\etal{et al. }
\newcommand\refer{\par \noindent\hangindent=3pc \hangafter=1}

\title{The Luminosity Function Evolution of Soft X--ray selected AGN in the
RIXOS survey }
\author[Page \etal]{M.\,J. Page\(^{1}\), F.\,J. Carrera\(^{1}\),
G. Hasinger\(^{2}\), K.\,O. Mason\(^{1}\), R. M\(^c\)Mahon\(^{3}\), \and
J.\,P.\,D. Mittaz\(^{1}\), X. Barcons\(^{4}\), R. Carballo\(^{4,5}\), 
I. Gonz\'alez--Serrano\(^{4}\),\and
 and I. P\'erez--Fournon\(^{6}\)\\
\(^{1}\)Mullard Space Science Laboratory, University College London, 
Holmbury St Mary, Dorking, Surrey RH5 6NT, UK.\\
\(^{2}\)Astrophysikalisches Institut Potsdam, An der Sternwarte 16,
Potsdam, Germany.\\
\(^{3}\)Institute of Astronomy, Madingley Road, Cambridge CB3 0HA,
UK.\\
\(^{4}\)Instituto de Fisica de Cantabria, CSIC--Universidad de Cantabria, 39005
Santander, Spain.\\
\(^{5}\)Dept. Fisica Moderna, Universidad de Cantabria, 39005 Santander,
Spain.\\
\(^{6}\)Instituto de Astrofisica de Canarias, Tenerife, Spain.}

\date{}

\begin{document}
\maketitle

\begin{abstract}

A sample of 198 soft X--ray selected active galactic 
nuclei (AGN)
from the {\it ROSAT} International X--ray Optical Survey (RIXOS), is used 
to investigate the X--ray
luminosity function and its evolution. RIXOS, with
a flux limit of  
\( 3 \times 10^{-14}{\rm erg\ s^{-1}\ cm^{-2}}\) (0.5 to 2.0 keV), samples a
broad range in redshift
over \({\rm  20\ deg^{2}}\) of sky, and is 
almost completely identified; it is used in combination with the {\it Einstein}
Extended Medium Sensitivity Survey (EMSS), to give a total sample of over
600 AGN.
We find the evolution of AGN with
redshift  to be consistent 
with pure luminosity evolution (PLE) models in
which the rate of evolution slows markedly or stops at high redshifts \(( z > 1.8 \)).
 We find that this result is not affected by the inclusion, or
exclusion, of
narrow emission line galaxies at low redshift in the RIXOS and EMSS samples,
and is insensitive to uncertainties in the conversion between
flux values measured with 
{\it ROSAT} and {\it Einstein}.
We confirm, using a model independent \(< V_{e}/V_{a} >\) test,
that our survey is consistent with no evolution at high redshifts.
\end{abstract}


\section{Introduction}

X--ray properties are becoming an increasingly important tool for selecting
samples of AGN, and sample sizes are approaching that of ultraviolet excess
(UVX) selected optical samples.

The largest currently available sample of serendipitous X--ray selected AGN
comes from the {\it Einstein} Extended
Medium Sensitivity Survey (EMSS, see Stocke \etal 1991), with a sky coverage
of \({\rm  778\ deg^{2}}\). However, the 
EMSS is dominated by low redshift objects (median {\it z} \(\sim\) 0.2) due to
its high limiting flux (typically \(> 10^{-13}{\rm erg\ s^{-1}\   
cm^{-2}}\) in the energy range 0.5 to 2.0 keV). Deeper {\it Einstein}
surveys (Primini \etal 1991) identified only 11 AGN. With the
arrival of {\it ROSAT} the possibilities for deeper surveys have 
been realised:
the Cambridge Cambridge {\it ROSAT} Serendipity Survey covers about 
\({\rm 4\  deg^{2}}\) and has a flux limit of 
\( 2 \times 10^{-14}{\rm erg\ s^{-1}\ cm^{-2}}\) from 0.5 to 2.0 keV (see
Boyle \etal 1995);
the survey of Boyle \etal (1994) contains 107 broad line AGN and 
probes to flux levels lower than 
\( 4 \times 10^{-15}{\rm erg\ s^{-1}\ cm^{-2}}\) (0.5 to 2.0 keV),
but is only 70\% complete at this flux limit and covers an area of less
than \({\rm  1.5\ deg^{2}}\) in total. This survey 
is almost devoid of low redshift objects, with only 3 AGN of 
\(z < 0.4\), and median
{\it z} = 1.5. 
Even deeper surveys, with even smaller sky coverage, 
(Hasinger \etal 1993; Branduardi--Raymont \etal 1994) 
are still in the process of optical
identification.

The {\it ROSAT} International X--ray Optical Survey (hereafter RIXOS, 
see Mason \etal 1995) occupies  a position between the EMSS and 
the deeper {\it ROSAT} surveys, with a sky coverage of over
\({\rm  20\ deg^{2}}\) and a limiting flux of 
\( 3 \times 10^{-14}{\rm erg\ s^{-1}\ cm^{-2}}\) (0.5 to 2.0 keV). 
 It is constructed from  81 long 
(\( > 8000\ {\rm seconds})\) {\it ROSAT}
pointings made with the Position Sensitive Proportional Counter (PSPC) at
the focus of the X--ray telescope. 
The RIXOS AGN sample consists of 198 objects, the largest {\it ROSAT}
selected AGN sample to date, and is particularly useful for 
investigating 
evolution models because of its broad sampling of redshifts. 

In this paper we present the results of an analysis 
 of the differential X--ray luminosity function (XLF) 
 and its evolution with redshift, using the RIXOS sample.
In section 2 we describe briefly the RIXOS survey 
and present details of the RIXOS AGN sample, while in section 3 we discuss the
log {\it N} -- log {\it S} of RIXOS AGN compared to that of the EMSS AGN. In section 4 the 
methods used
in this analysis are explained and in section 5 we present our
results, which are discussed in section 6. Our conclusions are presented 
in section 7.
 
Throughout this paper a Friedmann model universe 
has been assumed, and a value of  50 \({\rm km\ s^{-1}\ Mpc^{-1}}\) has been
adopted for the Hubble constant \( H_{0}\); two different values for the
deceleration parameter, \(q_{0}\) = 0 
and \(q_{0}\) = 0.5 have been used. 

\section{The RIXOS sample of AGN}

\subsection{Observations and Data Reduction}

The X--ray sources in RIXOS were found in a total of 81 {\it ROSAT} 
PSPC pointings. The target
of each observation and a small region around it have been excluded from the
analysis so that RIXOS 
consists entirely of serendipitously discovered sources. 
Sources more than 17
arcminutes offaxis have also been excluded due to their 
larger positional uncertainty 
and possible masking by the detector window support structure.
Furthermore, only sources detected in the harder {\it ROSAT} energy
band 0.4 to 2.4 keV are included; the poorer point spread function, 
interstellar absorption, diffuse Galactic
X--ray emission, and the increased contribution of Galactic stars 
 complicate the detection of extragalactic
sources in in the 0.1 to 0.4 keV band. 
Using finding charts from the automatic plate measuring (APM) facility 
	at the Royal Greenwich Observatory, Cambridge, optical spectra
were taken for all optical counterparts within the one-
sigma error circle of each X--ray source. If no likely counterpart was found,
 the optical counterparts in the larger two and three-sigma error 
circles were investigated. 
	Where the APM finding charts were not sufficient (for example if no
optical counterparts appeared near the X--ray source position) CCD images were
obtained using the Nordic Optical Telescope (NOT), or 
the Isaac Newton Group of telescopes  (ING) on La Palma. 
	Full details of the optical imaging and 
	spectroscopic observations and data
reduction are given in Mason \etal (1995).

	In this analysis, the term AGN is used to refer 
	to approximately the same
range of objects as in Maccacaro \etal (1991), that is objects with at least
one broad ( \(> 1000 {\rm km/s} \) ) emission line and/or 
[OIII]5007 \(>\) [OII]3727.
Hence the RIXOS AGN sample does include some narrow line objects. 
 These criteria have 
  been deliberately chosen to avoid any significant difference
between the RIXOS and EMSS optical selection, allowing 
the two samples to be meaningfully compared and combined. The effect of
excluding narrow line objects is discussed in section 6.

\subsection{Construction of the AGN sample used in this analysis}

\begin{table}
\caption{RIXOS \ cumulative sky coverage corrected for incompleteness}
\begin{tabular}{cccc}

     Flux Limit & Corrected & Optically & Number of \\
     \({\rm (erg\ s^{-1}\ cm^{-2})}\) & Area &
     Identified  & Fields \\
     \(0.5-2\) keV & \({\rm (deg^{2})}\) & Fraction & \\
\hline
                                          &  &  &               \\
     \({\rm 3.0 \times 10^{-14}}\) & 14.16 & 93\% & 62 \\
     \({\rm 3.5 \times 10^{-14}}\) & 14.36 & 95\% & 62 \\
     \({\rm 5.0 \times 10^{-14}}\) & 14.73 & 97\% & 62 \\
     \({\rm 6.0 \times 10^{-14}}\) & 15.09 & 99\% & 62 \\
     \({\rm 8.4 \times 10^{-14}}\) & 20.04 & 99\% & 81 \\

\end{tabular}
\end{table}

	A limiting X--ray flux of  \(3\times 10^{-14} 
{\rm erg\ s^{-1}\ cm^{-2}}\) (0.5 to 2.0 keV) was chosen for RIXOS, 
well above the detection
threshold for all the {\it ROSAT} fields used. Owing to the constraints of
optical telescope scheduling some X--ray sources remain unobserved and/or 
unidentified. In the interests of keeping incompleteness and optical selection
effects to a mimimum, the entire RIXOS survey has not been used.
In 62 of the 81 RIXOS  fields, all
objects have been identified or 
observed spectroscopically to the intended flux limit
of \({\rm 3.0 \times 10^{-14} erg\ s^{-1}\ cm^{-2}}\) (0.5 to 2.0 keV),
 while in each of the
 other
19 fields some, but not all, of the X--ray sources have been observed 
to this limit;
 these 
19 fields are, however, fully observed to a flux of \(8.4\times 10^{-14} 
{\rm erg\ s^{-1}\ cm^{-2}}\) (0.5 to 2.0 keV) and are included in the RIXOS AGN
sample with this flux limit. Overall spectroscopic completeness of the fields
used in the RIXOS AGN sample is 93\%; the remaining 7\% of sources which are 
unidentified are
those for which the optical counterpart(s) were too faint for us to obtain
reliable optical spectra. 
We have made the assumption that the fraction of 
unidentified sources which are AGN is the same as that for the
identified sources. Accordingly, the
sky area used for this analysis has been corrected,
 in a similar fashion to that of Boyle \etal (1994), by multiplying the area
by the fraction of sources identified; as the unidentified fraction is
small, this has only a small effect on our results. Since optical completeness 
is a
function of flux limit we have calculated the effective sky area at 
\({\rm 3.0 \times 10^{-14} erg\ s^{-1}\ cm^{-2}}\),
\(8.4\times 10^{-14} {\rm erg\ s^{-1}\ cm^{-2}}\), and three intermediate
fluxes corresponding to significant changes in spectroscopic completeness.
Again, the high level of completeness in RIXOS makes this a small
correction, which has only a small effect on our results. 
The number of {\it ROSAT} fields, 
corrected sky coverages and identified fractions 
at their respective limiting fluxes are listed in Table 1.

\begin{figure}
\begin{center}
\leavevmode
\epsfxsize=85mm
\epsfysize=80mm
\epsffile{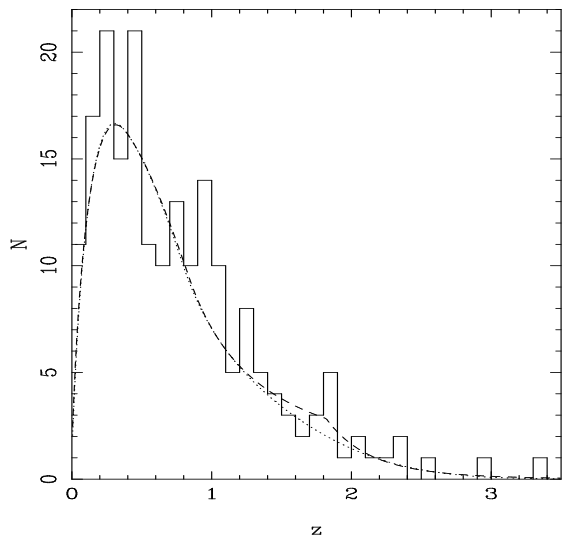}
\caption{Redshift Distribution, \(N(z)\), of RIXOS. The solid histogram is the
actual distribution while the dashed and dotted lines are the \(N(z)\) relations
predicted by the power law with redshift cutoff and polynomial models
respectively, for \(q_{0}\) = 0.}
\end{center}
\end{figure}

	The redshift distribution, \(N(z)\), 
of the RIXOS AGN sample is shown in Fig. 1. 
The sample has a significantly higher median redshift, 0.6, 
than the EMSS (0.2).

\begin{figure}
\begin{center}
\leavevmode
\epsfxsize=85mm
\epsfysize=100mm
\epsffile{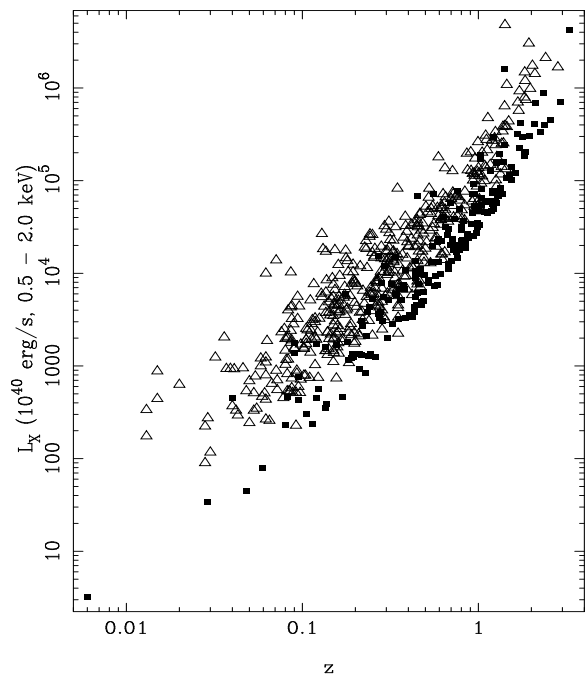}
\caption{X--ray Luminosity and redshift of RIXOS AGN (closed squares) and EMSS
AGN (open triangles). Luminosities have been calculated using \(q_{0}\) = 0.}
\end{center}
\end{figure}

	To obtain the largest possible working sample of AGN, 
the RIXOS and EMSS surveys
have been combined coherently (Avni \& Bancall 1980) 
to give a total of over 600 AGN. This combined sample will be referred to as 
`RIXOS + EMSS' hereafter. To correct
the  EMSS sample for incompleteness, the EMSS 
`expected' AGN (see Maccacaro \etal 1991) have been included in the EMSS and
EMSS + RIXOS samples. Throughout this analysis a power law X--ray spectrum has
been assumed, \(f_{\nu} \propto \nu^{-\alpha_{X}}\).
For comparison of the results presented in this paper with those of 
Boyle \etal (1994) and Maccacaro \etal (1991),
the analysis has been performed using \(\alpha_{X}\) = 1.
The first results from X--ray colour analysis  
(see Mittaz \etal 1995) indicate that this is
quite representative of the RIXOS AGN, which have a median \(\alpha_{X}\)
\(\sim\)
1.01. The scatter of \(\alpha_{X}\) around the value 1.0 appears to be
larger in the RIXOS sample than was found by Macaccaro \etal (1988) for EMSS
AGN, although the hardening of source spectra with redshift (see Francis 1993)
does not appear to be significant in the RIXOS sample and has not been
included
in this analysis.
The
X--ray luminosity - redshift (\( L_{X},z\)) distribution for the RIXOS and
EMSS AGN are compared in Fig. 2. As expected from a deeper survey, the RIXOS AGN
typically have lower luminosity and/or higher redshift than the EMSS AGN.  

\section{log {\it N} -- log {\it S}}

\begin{figure}
\begin{center}
\leavevmode
\epsfxsize=85mm
\epsfysize=110mm
\epsffile{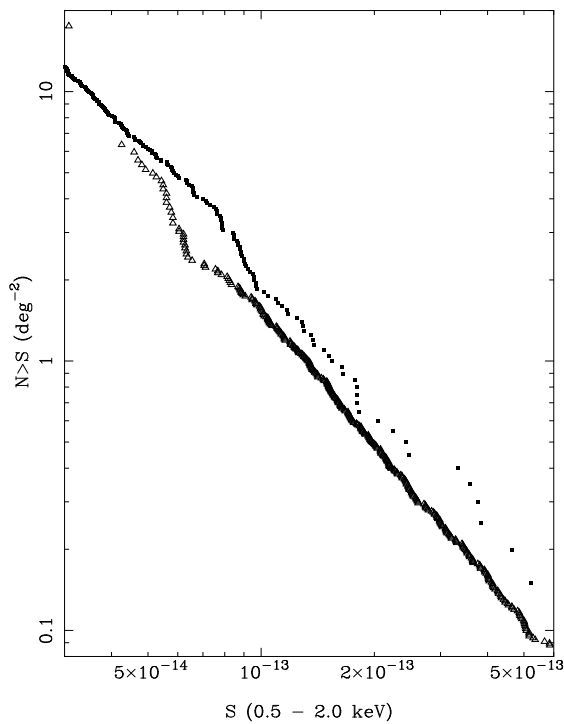}
\caption{Integral log {\it N} -- log {\it S} of RIXOS AGN (closed squares) and EMSS
AGN (open triangles).}
\end{center}
\end{figure}

The survey of Boyle \etal (1994) contains a larger number of AGN than would be
predicted by extrapolating the EMSS log {\it N} -- log {\it S} to lower fluxes; 
these authors suggested that this may be partly due to an error in their 
conversion from {\it ROSAT} to {\it Einstein} fluxes.  In practice there are a
number of factors which might affect the relative source counts and/or fluxes
of the two samples, including differences in the source detection and
parameterisation algorithms, uncertainties in the detector response matrices
used to transform count rates to fluxes, uncertainties in the spectral form 
used to convert from fluxes in the  {\it Einstein} 0.3 to 3.5 keV band to the
{\it ROSAT} 0.5 to 2.0 keV band and the effects of incompleteness or other
biases. 

To assess these effects  Figure 3 compares  the integral log {\it N} -- log
{\it S} for the RIXOS and EMSS AGN. This plot is derived using a conversion
factor (CF) of 1.8 between the fluxes in the two bands which  is appropriate
for a powerlaw spectrum of $\alpha_x = 1$ if we use the standard published
response matrices for {\it ROSAT} and {\it Einstein}.  It is clear that RIXOS
has a larger number of sources  than the EMSS  by about 30\%  at
almost all fluxes when using this conversion. Although RIXOS samples a larger
area than the EMSS at low fluxes, there is significant overlap in the fluxes of
objects found in the two samples, and there is thus  no physical reason why the
log {\it N} -- log {\it S} relations of the two surveys should be different.

Further investigation of the reasons for  this discrepancy is beyond the scope
of this paper.  Instead we adopt an approach whereby we parameterise the
discrepancy empirically, and investigate to what extent the uncertainty 
in this number affects the
results on AGN evolution when we combine the RIXOS and EMSS samples. The device
we use for this empirical parameterisation is to find the CF for
which
the two  log {\it N} -- log {\it S}
relations are consistent. At the flux limits of the EMSS and RIXOS the 
{\it N(S)} relation is well fit by a power law.  
Using maximum likelihood, the two
samples have been fitted simultaneously with a single power law slope but
different normalisations. 
\[\frac{dN}{dS}=k_{E}S^{-\gamma} \ \ {\rm EMSS \
objects,\ S(0.3 - 3.5keV)}\] 
\[\frac{dN}{dS}=k_{R}S^{-\gamma} \ \ {\rm RIXOS \
objects,\ S(0.5 - 2.0keV)}\] 
There are two free parameters in this fit, the
power law slope \(\gamma\)  and the difference between the two normalisations
(\(k_{E}/k_{R}\)); the actual normalisations  are found by requiring that the
number of objects predicted by the  log {\it N} -- log {\it S} for the sky
coverage of RIXOS plus the sky coverage of the EMSS is equal to the total in
RIXOS plus the total in the EMSS. The normalisation difference should be
related simply to the empirical CF from {\it ROSAT} to {\it Einstein} 
fluxes by  
\[{\rm CF} = (k_{E}/k_{R})^{1/(\gamma -1)}\] 
We assumed upper flux limits of \(2\times 10^{-11} 
{\rm erg\ s^{-1}\ cm^{-2}}\) (0.3 to 3.5 keV) and \(10^{-12} 
{\rm erg\ s^{-1}\ cm^{-2}}\) (0.5 to 2.0 keV) for the EMSS and RIXOS
respectively, to reflect the selection against very bright sources in these
surveys. This has only a small effect on the results; changing the upper
flux limits to any reasonable value (or removing them) changes the best fit by
only a fraction of the 1 \(\sigma\) statistical errors quoted below.
The best fit slope
\(\gamma\) is found to be 2.56, consistent with the  slopes found when the two
samples are fit independently,  \(2.61 \pm 0.06\) for the EMSS and
\(2.45 \pm 0.11\)  for RIXOS, where errors are 1 \(\sigma\). 
Since the slope of the EMSS
log {\it N} -- log {\it S} curve is actually 
steeper than (although consistent with) 
that of RIXOS, there is no evidence from this comparison 
to support 
previous claims about  incompleteness in the EMSS 
(e.g. Franceschini \etal 1994) in which incompleteness is thought to be a
problem at low fluxes. 
The best fit CF
is found to be \(1.47 \pm 0.11\); again errors are 1 \(\sigma\).  The 
standard conversion factor of 
1.8  is thus  significantly different (rejected at \(>\)99.0\%)  from that
found by matching the log {\it N} - log {\it S} curves.  There is, however,  no
evidence from the RIXOS and EMSS log {\it N} - log {\it S}  that the CF should
be as small as 1.0, a possibility considered in Boyle \etal (1994).
                              
To assess the impact of the different conversion factors on AGN evolutionary
models, values of 1.47 and 1.8 have both been used in the subsequent analysis
and the results compared. Note that all fluxes and luminosities quoted in
this paper are 0.5 to 2.0 keV, i.e.  the  {\it Einstein} fluxes have been
converted to the {\it ROSAT} flux band.

\section{The Luminosity Function}

The techniques that we have applied to the data are discussed in this
section, while the results of the analysis are described in section 5.

A flux limited survey samples different luminosity ranges at different
redshifts.
	The AGN population as a function of luminosity and redshift is
represented by 
 the 
XLF, \(\phi (L,z)\), 
which is defined as the number of objects detected per unit comoving volume
per unit luminosity interval, i.e.
           \[\phi (L,z)=\frac{d^{2}N}{dV dL}(L,z)\]

Evolution in the space density or luminosity of AGN with redshift is seen as a
change in the XLF.

The two simplest forms for evolution are pure density and pure luminosity
evolution (PLE). In pure density evolution, the number of AGN per unit 
comoving volume is
assumed to evolve while the distribution of luminosities remains constant.
In PLE, the number of AGN per unit comoving volume remains
constant, while the luminosity of each AGN evolves. 

Pure density evolution models do not predict the flattening at low
fluxes which is seen in 
the best log {\it N} -- log {\it S} curves currently
available, 
in both the optical (Boyle \etal 1988) and X--ray bands (Branduardi -
Raymont \etal 1994, Hasinger \etal 1993). For this reason, recent 
 evolutionary
models have been based on luminosity evolution of some form.
	Only PLE models are considered in this paper.

As shown by Maccacaro \etal (1991), the XLF can be  modeled as two power laws 
with a break luminosity, 
\( L_{break}\), such that
          \[\phi= K_{1}L^{-\gamma_{1}} \ \ L < L_{break}\]
          \[\phi= K_{2}L^{-\gamma_{2}} \ \ L > L_{break}\]
            
where \( K_{1}\ and \ K_{2}\ \) are normalisations of the two power laws. Since
we 
require the luminosity function to be continuous, the two normalisations are
not independent. A single normalisation 
\( K_{1}\)  is adequate, since \[K_{2} = K_{1}/L_{break}^{(\gamma_{1}-\gamma_{2})}\]

	In the framework of a PLE model the XLF retains its shape
at all redshifts, hence the XLF at any redshift depends on 
only the evolution law and the XLF at
zero redshift (hereafter {\it z} = 0 XLF).

\subsection{Maximum likelihood and 2D K.S. testing}

The maximum likelihood method (Marshall 1984) has been used to obtain
best fit evolution parameter(s) and {\it z} = 0 XLF for each evolution model,
utilising the full RIXOS + EMSS sample of AGN. This technique involves simultaneously
fitting the evolution and the {\it z} = 0 XLF. There are four or five
free parameters (three
from the {\it z} = 0 XLF, \( \gamma_{1},\ \gamma_{2}\), and \(L_{break}\), 
plus either one or two 
from the evolution model) in fitting the models considered in this
paper. The normalisation of the {\it z} = 0 XLF is
set so that the total number of 
objects predicted by the model is equal to the number
in the sample, and is not a free parameter in the fit. 

To test the acceptability of evolutionary models, the two
dimensional Kolmogorov-Smirnof (2D K.S.) test  has
been used. Two alternatives have been used in the literature, 
the test described by
Peacock (1983), and the test of Fasano \&
Franceschini (1987), which is also described in Press \etal (1992). 
It is important to use the test most effective at
rejecting poor models and accepting good models for the RIXOS and EMSS
datasets, to ensure confidence in the
results.
To assess the most appropriate test to use, both were   
 applied to simulated data as follows:
    
 The two best fit XLFs and evolution models (see section 5.1) 
 from Maccacaro \etal (1991) 
 were used to
produce 200 simulated samples of AGN, 100 from each evolution model. Each
sample was constructed using two completeness limits, appropriate to the
RIXOS and EMSS samples respectively. The number of objects in each
simulated sample 
was between 400 and 600, similar to the number of objects in the combined
RIXOS + EMSS sample, and the tests were performed over exactly the same
plane as those used on the real sample (see below). In this way, the 2D K.S.
tests were evaluated under very similar conditions to those under which they
were actually to be applied. 
Each sample was tested once against its own parent XLF
and evolutionary model, and once against the other, so that 200 tests were
performed with correct models and 200 with incorrect models. The Peacock
test rejected 100 incorrect models (50\%) at the 95\% level and 6 correct 
models, while the Fasano \& Franceschini test rejected 84 incorrect
models (42\%) and 10 correct models. 
This indicates that the 2D K.S. test of Peacock is more
efficient at distinguising between good and bad models of our data. 
Because the EMSS and RIXOS contain a very large range
of sky coverage at different completeness limits, the combined RIXOS + EMSS
sample has a slightly lower redshift - de-evolved luminosity correlation 
coefficient (typically 0.43, but dependent on cosmological and evolutionary
model) 
than the simulation models (typically 0.5 to 0.7). According to Fasano
\& Franceschini (1987), their test  reaches maximum efficiency when compared to
that of Peacock (1983) at {\it higher} correlation coefficients, and so it is
reasonable to assume the results obtained from our simulations
should hold for the actual data to be tested. The test of Peacock (1983)  
has therefore been used in this paper.

Models have been tested in the redshift and 
de-evolved luminosity plane over a range that includes all parameter values
found in RIXOS and the EMSS, 
 \( (0 < z < 3.5, 10^{40} < L_{0} <  10^{47})\) where 
\( L_{0}\) is the de-evolved 0.5 - 2.0 keV luminosity in 
\({\rm erg\ s^{-1}}\);
note that \( L_{0}\) is model dependent.  Since there is 
no selection criterion in this analysis, based on observed X--ray
luminosity, which would correspond to the \(M_{B} < -23\) requirement often used in
optical QSO surveys, the test has
not been performed over an interval in  {\it observed} 
luminosity. Imposing a lower limit to   
observed luminosity in this way would introduce implicit
density evolution to a PLE model (see Kassiola and Mathez 1990), and is
hence undesirable.

\subsection{\(1/V_{a}\) and \(< V_{e}/V_{a} >\)}

An estimate of the behaviour of the XLF can be gained in a
model-independent way using the \(1/V_{a}\) 
statistic (Avni \& Bahcall 1980) and plotting the XLF
 in distinct redshift bins, as in Maccacaro \etal (1991). Of course, any
 evolution occuring within these distinct bins will not be apparent in this
type of treatment. 

A more quantitative treatment can be made using the \(< V_{e}/V_{a} >\) 
statistic (Avni \& Bahcall 1980).
To investigate single parameter PLE models, 
without simultaneously
 modelling the {\it z} = 0 XLF, we have carried out 
\(<V_{e}/V_{a}>\) tests in distinct redshift shells.
In this case the test is used to obtain acceptable evolution parameters for
the given model, and  
the bins have been chosen so that within each shell the evolution parameter has
a 68\% confidence region of about \(  \pm  20\% \) its value over the entire
redshift range; this is a good compromise between resolution in redshift and
constraint of the evolutionary properties of each bin.
 The practical aspects of the \(< V_{e}/V_{a} >\) test within redshift shells
are
described in Della Ceca \etal (1992).

We have also used the \(< V_{e}/V_{a} >\) test 
to examine evolution at high redshift in a model independent way. In this case 
the \(< V_{e}/V_{a} >\) test has been applied in the redshift interval 
\(z_{b} < z < 3.5\)
where \( z_{b}\) is varied. Here, the test is used with no
evolutionary model, and is capable of determining whether the data are
consistent with the no evolution hypothesis between \(z_{b} < z < 3.5\),
and if not, whether the luminosity function is increasing with redshift 
(\(<V_{e}/V_{a}>\ \  > 0.5\)), or decreasing with redshift 
(\(<V_{e}/V_{a}>\ \  < 0.5\)).
This use of the \(< V_{e}/V_{a} >\) test is described in 
Dunlop and Peacock (1990).

\section{Results}

\begin{table*}
\caption{Results of fitting evolution models}
\begin{minipage}[t]{10in}
\begin{tabular}{cccccccccccccc}
\small

 model&CF&
\(q_{0}\) 
&\( z_{cut}\)&C&\(C_{1}\)& 
\(K_{1}\)\footnote{\(K_{1}\) in units of  
\({\rm 10^{-4}(10^{40}erg s^{-1})^{(\gamma_{1}-1)}\ Mpc^{-3}}\)}& 
\( \gamma_{1}\) & \(\gamma_{2}\) & \(log_{10}\) 
& \(I_{XRB}\)\footnote{\(I_{XRB}\) in units of \({\rm 10^{-9}\ erg\
s^{-1}\ cm^{-2} sr^{-1}}\) (1 - 2 keV)}
& P(\(>\)D) &
P(\(>\)D)&P(\(>\)D)\\
      &   &  &  &  &  &   &
   &   &  \((L_{break})\footnote{\(L_{break}\) in units of \({\rm
10^{40}erg\ s^{-1}}\)}\) & &RIXOS&RIXOS&EMSS\\ 
      &   &   &  &   &   &  &  &   &   &  &+EMSS&  & \\
\hline
      &   &   &  &   &   &   & &  &   &  &  &  &   \\
\((1+z)^{C}\)& 1.8 & 0.0 &  -- 
& 2.66 & -- & 2.18 
& 1.68 & 3.38 & 3.59 & 8.24 & 0.020 & 0.26 & 0.036 \\
\(e^{C\tau}\)& 1.8 & 0.0 & -- & 4.74 & -- & 
1.59 & 1.64 & 3.23 & 3.29 & 7.31 & 0.033 & 0.047 & 0.033 \\
\((1+z)^{C}\)& 1.8 & 0.0 & 1.82 & 2.91 & -- & 
1.83 & 1.65 & 3.30 & 3.49 & 6.88 & 0.27 & 0.62 & 0.096 \\
\(10^{Cz+C_{1}z^{2}}\)& 1.8 & 0.0 & -- & 1.10 & \(-0.230\) & 
1.56 & 1.62 & 3.27 & 3.52 & 6.47 & 0.36 & 0.45 & 0.17 \\
      &   &   &    &   &   &   &   & &   &  &  &  &   \\
\((1+z)^{C}\)& 1.8 & 0.5 & -- 
& 2.35 & -- & 1.75 
& 1.62 & 3.38 & 3.57 & 4.56 & 0.015 & 0.62 & 0.017 \\
\(e^{C\tau}\)& 1.8 & 0.5 & -- & 3.82 & -- & 
1.37 & 1.63 & 3.32 & 3.20 & 4.71 & 0.019 & 0.21 & 0.0048 \\
\((1+z)^{C}\)& 1.8 & 0.5 & 1.42 & 2.94 & -- & 
1.22 & 1.57 & 3.30 & 3.38 & 4.18 & 0.69 & 0.24 & 0.25 \\
\(10^{Cz+C_{1}z^{2}}\)& 1.8 & 0.5  & -- & 1.07 & \(-0.238\) & 
1.48 & 1.60 & 3.33 & 3.46 & 4.20 & 0.58 & 0.58 & 0.21 \\
      &   &   &   &   &   &   &   & &   &  &  &  &   \\
\((1+z)^{C}\)& 1.47 & 0.0  & -- 
& 2.46 & -- & 2.00
& 1.64 & 3.23 & 3.64 & 8.28 & 0.055 & 0.55 & 0.042 \\
\(e^{C\tau}\)& 1.47 & 0.0  & -- & 4.44 & -- & 
1.81 & 1.65 & 3.16 & 3.40 & 7.71 & 0.016 & 0.095 & 0.064 \\
\((1+z)^{C}\)& 1.47 & 0.0 & 1.82 & 2.72 & -- & 
2.19 & 1.66 & 3.23 & 3.60 & 7.35 & 0.22 & 0.86 & 0.19 \\
\(10^{Cz+C_{1}z^{2}}\)& 1.47 & 0.0 & -- & 1.04 & \(-0.219\) & 
1.92 & 1.64 & 3.20 & 3.58 & 7.03 & 0.24 & 0.65 & 0.20 \\
      &   &   &   &   &   &   &   & &  &  &  &  &   \\
\((1+z)^{C}\)& 1.47 & 0.5 & -- 
& 2.19 & -- & 1.64 
& 1.59 & 3.23 & 3.62 & 4.88 & 0.049 & 1.2 & 0.019 \\
\(e^{C\tau}\)& 1.47 & 0.5 & -- & 3.47 & -- & 
1.52  & 1.61 & 3.19 & 3.31 & 4.86 & 0.026 & 0.24 & 0.045 \\
\((1+z)^{C}\)& 1.47 & 0.5 & 1.41 & 2.73 & -- & 
1.39 & 1.57 & 3.19 & 3.48 & 4.57 & 0.22 & 0.38 & 0.14 \\
\(10^{Cz+C_{1}z^{2}}\)& 1.47 & 0.5 & -- & 1.04 & \(-0.238\) & 
1.54 & 1.58 & 3.22 & 3.51 & 4.69 & 0.38 & 0.80 & 0.18 \\
\hline
\( \pm \)ERRORS &   &   &   &   &  & &   &   &   &  &  &  &   \\
\((1+z)^{C}\)&   & & 
& 0.08 &  & 
& 0.08 & 0.08 & 0.10 &  &  &  & \\
\(e^{C\tau}\)&  & & & 0.20 & & 
 & 0.07 & 0.08 & 0.05 &  &  &  & \\
\((1+z)^{C}\)& & & 0.10 & 0.10 & & 
 & 0.07 & 0.08 & 0.05 &  &  & & \\
\(10^{Cz+C_{1}z^{2}}\)&  &  & & 0.06 & 0.03 & 
 & 0.10 & 0.09 & 0.06 & & &  & \\
\hline
\end{tabular}
\end{minipage}
\end{table*}

	The best fit evolution parameters and luminosity functions have been
obtained from the
combined RIXOS + EMSS sample. As a check for consistency
between the two samples they have been tested for goodness of fit both 
individually
and in combination.
	Table 2 shows the results for the maximum likelihood and 
2D K.S. tests applied to the models that have been investigated. Errors
quoted were obtained using the method of Lampton, Margon and Bowyer
(1976), and correspond to \(\Delta \chi^{2}\) = 1, i.e. 68\% confidence
intervals for one interesting parameter.

\subsection{The simplest models}

The two simple PLE models used by Maccacaro \etal (1991), are power law evolution
\[L = L_{0} \times (1+z)^{C}\]
and evolution which is exponential with look back time
\[L = L_{0} \times e^{C \tau}\]
where  C is the evolution parameter and \(\tau\) is look back time.
	\[\tau=z/(1+z) \ \ q_{0} = 0\]
	\[\tau=1-1/(1+z)^{3/2} \ \ q_{0} = 0.5\]
The 2D K.S. test rejects both these simple models at the 95\% level for the
combined sample except the \(q_{0}=0\), CF = 1.47 case where
the power law model is just acceptable at the 95\% level. 
Note that the case where the
2D K.S. probability P\((>D)\) is 1.2 indicates that the difference between the expected and actual
distributions is small, not that it is zero;
it is possible for the test
of Peacock (1983) to produce values for the P\((>D)\) larger 
than 1, 
in which case it cannot be regarded as a probability,
although the implication that a model is a good fit if
P\((>D)\) is high is certainly true. 

\begin{figure}
\begin{center}
\leavevmode
\epsfxsize=85mm
\epsfysize=90mm
\epsffile{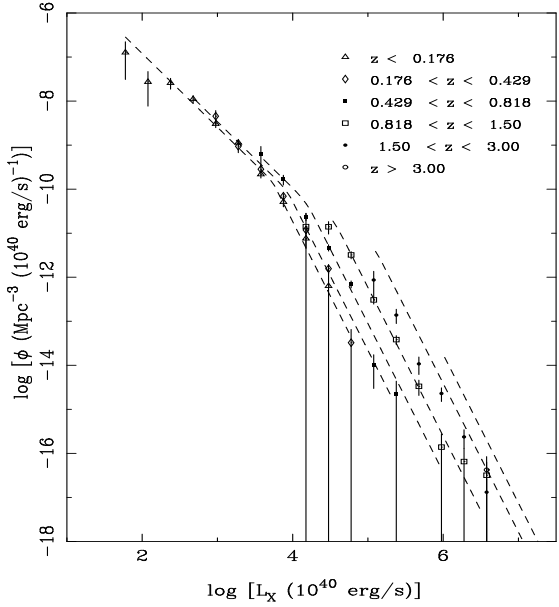}
\caption{Binned \(<1/V_{a}>\) XLF of the RIXOS + EMSS sample and best fit 
 power law evolution model XLF (dashed lines) for  \(q_{0}\) = 0 and CF = 1.8.}
\end{center}
\end{figure}

\begin{figure}
\begin{center}
\leavevmode
\epsfxsize=85mm
\epsfysize=90mm
\epsffile{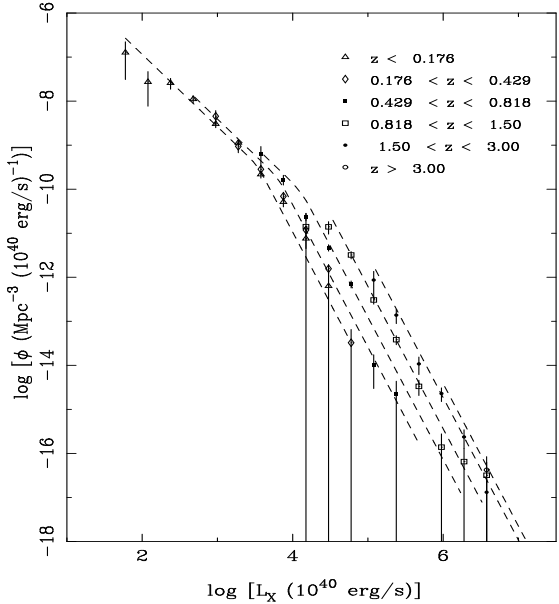}
\caption{Binned \(<1/V_{a}>\) XLF of the RIXOS + EMSS sample and best fit 
 exponential evolution model XLF (dashed lines)  
 for  \(q_{0}\) = 0 and CF = 1.8.}
\end{center}
\end{figure}

Figs 4 and 5 show binned \(1/V_{a}\) luminosity functions in
redshift shells against the best fit power law and exponential evolution model
luminosity functions. Figs 6 and 7 
shows evolution parameters acceptable at 68\%
for these two evolution models derived from the \(< V_{e}/V_{a} >\) test in
redshift intervals. These figures suggest that the exponential model
fails because it requires unacceptably rapid evolution at low redshift, (in
Fig. 7 the low redshift evolution parameter lies below any value which
would be consistent with \(z > 0.4\)), while
the power law model fails because it overpredicts evolution at high redshift, 
(for \( z > 1.5\) the model curves in Fig. 4 lie well above the data).
Evolution is slower at high redshift in the exponential model than in the power
law model, and hence in Fig. 5 the exponential model appears less discrepant
at \(z > 1.5\) than the power law model in Fig. 4. 
Figs 4 to 7 have been constructed using
\(q_{0}\) = 0, CF = 1.8.

\begin{figure}
\begin{center}
\leavevmode
\epsfxsize=85mm
\epsfysize=70mm
\epsffile{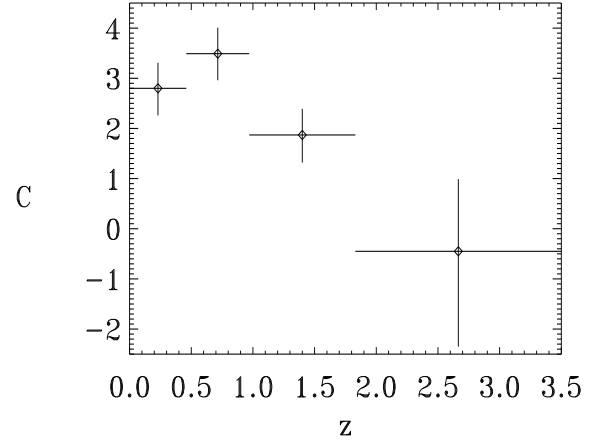}
\caption{Power law evolution parameter C in redshift bins using the 
\(< V_{e}/V_{a} >\) test  for  \(q_{0}\) = 0 and CF = 1.8.}
\end{center}
\end{figure}

\begin{figure}
\begin{center}
\leavevmode
\epsfxsize=85mm
\epsfysize=70mm
\epsffile{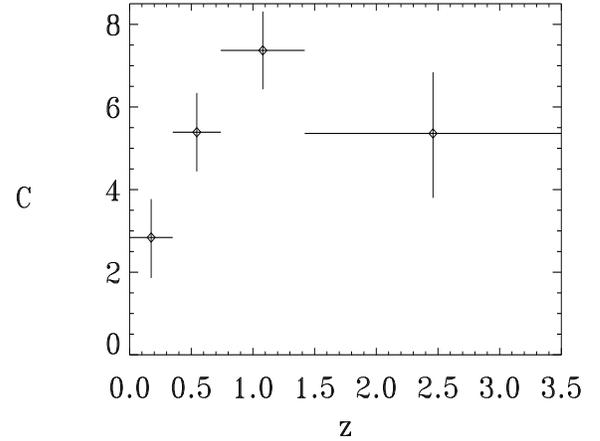}
\caption{Exponential evolution parameter C in redshift bins using the 
\(< V_{e}/V_{a} >\) test  for  \(q_{0}\) = 0 and CF = 1.8.}
\end{center}
\end{figure}

\subsection{Two Parameter Evolution Models}

As both single parameter models are rejected by the 2D K.S. test, more complex
models have been investigated. From Figs 4 and 5 it appears there is
 little difference in the
slope of the luminosity function in different redshift bins, so models in which the
luminosity function changes slope with redshift have not been considered. 
A power law model with
a redshift cutoff, \(z_{cut}\), where
\[L = L_{0} \times (1+z)^{C} \ \ z < z_{cut}\]
\[L = L_{0} \times (1+z_{cut})^{C} \ \ z > z_{cut}\] 
and alternatively, a polynomial evolution of the
form  \[L = L_{0} \times 10^{(Cz+C_{1}z^{2})}\] both have two free
parameters for the evolution (plus three for the luminosity function). Both of
these models are accepted 
at the \ 95\% \ level by the 2D K.S. test for the
combined RIXOS + EMSS samples for both values of \( q_{0}\) and CF; there is
little justification, from the 2D K.S. probabilities obtained, to prefer one
of the models.  
 The similarity of the two models for \(z<1.5\)
 is illustrated by the predicted 
\(N(z)\) relations plotted in Fig. 1. The two evolution models are however
radically different in shape beyond this redshift; in the polynomial evolution
model luminosity declines after  {\it z} \(\sim\) 2, and at {\it z} = 3.5 
the expected number
of objects differs by a factor of 5 for the two models. A larger sample of 
{\it z} \(>\) 2
objects would have the potential to discriminate between the two models. 

\subsection{Evolution at high redshift}

It is seen in the previous section that 
PLE models in which evolution ceases or changes direction at
high redshift are found more acceptable than models where evolution continues, 
for both values of \(q_{0}\)  and CF, 
indicating that evolution at {\it z} \(>\) 2
must be absent or slow compared to that at low redshift.
As further evidence, 
the  results of the \(< V_{e}/V_{a} >\) test from {\it z} = \( z_{b}\)
to {\it z} = 3.5 for the combined RIXOS + EMSS 
sample are shown in Figs 8 and 9.
Both were constructed using a CF of 1.8; when 1.47 is used,
\(<V_{e}/V_{a}>\) is a few percent lower; note that above \( z_{b}\) = 2.4 
there are only 4 objects included in the test. 
From \( z_{b}\) = 1.67, with 30 AGN, 
the \(<V_{e}/V_{a}>\) test shows the data to be
consistent at the 
68\% level 
with no evolution for both values of \(q_{0}\) and CF.
The \(<V_{e}/V_{a}>\) test used in this way 
is model independent, and this result is {\it not}
restricted to PLE models.

\section{Discussion}

\begin{figure}
\begin{center}
\leavevmode
\epsfxsize=85mm
\epsfysize=80mm
\epsffile{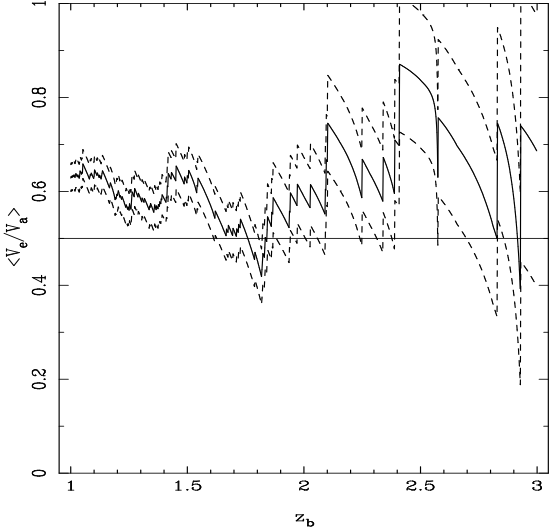}
\caption{\(< V_{e}/V_{a} >\) test in the redshift interval {\it z} = \(z_{b}\)
to  {\it z} = 3.5 using the RIXOS + EMSS sample for \( q_{0}\) = 0.}
\end{center}
\end{figure}
 
\begin{figure}
\begin{center}
\leavevmode
\epsfxsize=85mm
\epsfysize=80mm
\epsffile{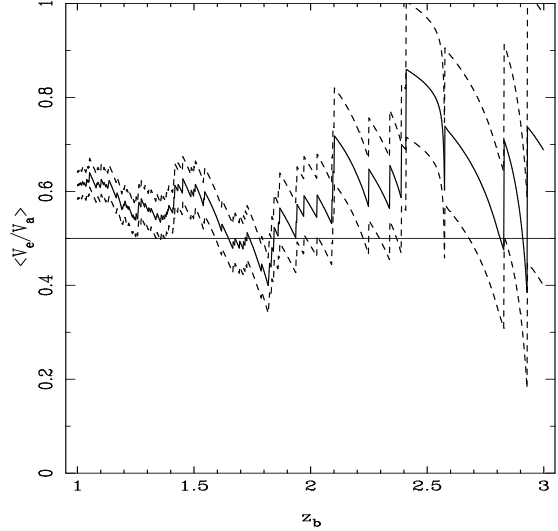}
\caption{\(< V_{e}/V_{a} >\) test in the redshift interval {\it z} = \(z_{b}\)
to  {\it z} = 3.5 using the RIXOS + EMSS sample for \( q_{0}\) = 0.5.}
\end{center}
\end{figure}

Although changing the {\it ROSAT} to {\it Einstein} CF from 1.8 to 1.47
does not appear to affect the choice of model, (i.e. the power
law with evolution cutoff and polynomial models are both significantly
more acceptable than the simple power law and exponential models), it does have a
significant effect on the best fit evolution parameter(s)
and the {\it z} = 0 XLF. In all cases the best fit parameters for CF = 1.47
are outside the 90\% confidence region of the parameters for CF = 1.8;
evolution is slower, and the steep part of the XLF is less steep
if the CF of 1.47 is used. The slope of the low luminosity region of the XLF
is
affected little by the choice of CF.

The evolution parameter, \(C\), for the
exponential model is consistent at 68\% with the EMSS value
(\(C\ =\ 4.18 \pm 0.35\) for \(q_{0}\) = 0) only for CF = 1.47, and its low K.S.
probability confirms the result of Della Ceca \etal (1991) that this  model
is a poor description of AGN evolution.
For the power law model, the value of \(C\) obtained here is consistent
with that found by Maccacaro \etal (1990) (\(2.56 \pm 0.17\) for \(q_{0}\)
= 0) for both values of CF. 
The introduction of a cutoff in evolution at \(z\)
= 1.8 significantly improves the 2D K.S. probability of the power law
model for the RIXOS + EMSS sample; the last column of Table 2 shows that the
EMSS data alone is also better fit with the evolution cutoff. Including
an evolution cutoff at \(z\) = 1.8 in  
a \(<V_{e}/V_{a}>\) test to the EMSS data with a power law evolution model
and \(q_{0}\) = 0 
gives \(C\) = \(2.74\pm 0.20\), almost identical to the  
RIXOS+EMSS value for CF = 1.47 and still consistent for CF = 1.8.

Comparing the CF = 1.8, \(q_{0}\) = 0 evolution parameters and {\it z} = 0
XLFs
from the RIXOS+EMSS sample
with those of Boyle \etal (1994), we find similar values for all but the
exponential evolution model which evolves faster
in Boyle \etal (1994). For \(q_{0}\) = 0.5, there is less agreement, with
the {\it z} = 0 XLFs of the exponential, power law, and polynomial evolution
laws having significantly lower values for \(\gamma_{1}\) (i.e. flatter
slopes at low luminosity) in Boyle \etal (1994). The best fit \(q_{0}\) = 0.5
power law with
evolution cutoff model in Boyle \etal (1994), whilst having a similar 
{\it z} = 0 XLF and evolution rate \(C\), has a much
higher cutoff redshift (\(z_{cut}\) = 1.7), than the value found here, 
\(z_{cut}\) = 1.4 for \(q_{0}\) = 0.5.

It is notable that almost all of the models considered in this paper are
found more acceptable to the 2D K.S. test than in Boyle \etal (1994), in
which almost all models are rejected at \(>\) 99\%.
 Boyle \etal used the test of Fasano \& Franceschini (1987), while we have
used that of Peacock (1983). However we have also
tested our models against the RIXOS +
EMSS dataset using the test of Fasano \& Franceschini, and do not find 
them to be rejected at \(>\) 99\%.
In Boyle \etal (1994), the inclusion of narrow line objects gave
significantly increased 2D K.S. probabilities, (i.e. a better fit).
However, narrow line
objects which would be
classified by Stocke \etal as AGN have been included throughout this analysis.
If we exclude these objects, which occurr in both the EMSS and RIXOS, we do
find
lower 2D K.S. probabilities, although the polynomial and power law with
evolution cutoff models are still acceptable at the 95\% level while the
simple 
power law
and exponential evolution models are not. For the {\it z} = 0 XLF, typically
\(\gamma_{1}\) is reduced by about 0.1 and \(L_{break}\) is increased by
25\% to 50\% depending on the specific PLE model and choice of cosmology;
the best fit values of \(\gamma_{2}\) and \(z_{cut}\) change
by no more than 0.02 and the evolution parameters remain within the errors
quoted in table 2.
There are no broad line objects with \(L_{0} < 2 \times
10^{41} {\rm erg\ s^{-1}}\).

The poor model fits of Boyle \etal (1994) may be attributable to the
conversion between {\it ROSAT} and {\it Einstein} measured fluxes;
the {\it ROSAT} sample of Boyle \etal (1994)
combined with the  EMSS AGN using a CF of 1.8
shows an evolution rate too high to be
consistent with the EMSS sample alone. 
 As
we have discussed in section 3, the effective 
CF from {\it ROSAT} to {\it EINSTEIN}
fluxes may be significantly lower than 1.8, and as seen in table 2 a lower
CF has the potential to reduce the evolution rate of combined 
{\it ROSAT} and {\it EINSTEIN} data, and hence could 
improve the self consistency
and model fits of Boyle \etal (1994).

The log {\it N} - log {\it S} at faint fluxes derived from the CF = 1.8 models
tested above, and extended to {\it z} = 4 are shown in Fig. 10. 
None of the model curves exceed the
total faint X--ray 
log {\it N} - log {\it S} obtained by fluctuation analysis in Hasinger
\etal (1993), or  Barcons \etal (1994). Notably, the log {\it N} - log {\it S}
curves at faint fluxes for PLE are 
separated strongly by the value of \(q_{0}\) used, while the
specific choice of PLE model has a comparatively minor effect. 
In a \(q_{0}\) = 0.5
universe, AGN undergoing PLE should represent between 30\% and 45\% 
of all sources with 
\(S > 10^{-15} {\rm erg\ s^{-1}\ cm^{-2}}\); in contrast, in a \(q_{0}\) = 0
universe AGN evolving in this way would be expected to 
constitute between 55\% and 100\% of these
sources. Above 
\(3\times 10^{-14} {\rm erg/s/cm^{2}}\) the log {\it N} - log {\it S} curves
from the PLE models are all very similar and represent the data well.

The AGN contribution to the \(1-2\) keV X--ray background has been
calculated for \(0<z<4\), \(10^{40}<L_{0}<10^{47}\) (where \(L_{0}\) is
the de-evolved 0.5 - 2 keV luminosity in erg \({\rm s^{-1}}\))
for all the models tested in section 5 and is shown 
in table 2 (column entitled \(I_{XRB}\)). As expected from the log 
{\it N} - log {\it S} predictions, the contribution of AGN to the 
X--ray background
from these models has a stronger dependence on the value of
\(q_{0}\) than the choice of PLE model. The values for the AGN X--ray
background intensity given in table 2 are in good
agreement with those of Boyle \etal (1994).
A recent measurement of the X--ray background (Chen, Fabian and Gendreau 1996)
using both {\it ASCA} and {\it ROSAT} found an intensity of 1.46 \({\rm 
\times 10^{-8} erg s^{-1} cm^{-2} sr^{-1}}\) (1 - 2 keV). Using this value, our
acceptable models (power law with evolution cutoff and polynomial) predict
that AGN account for between 44\% and 50\% (\(q_{0}=0\)) or 29\% and 32\%
(\(q_{0}=0.5\)) of
the 1 - 2 keV X--ray background, where these ranges include the uncertainty
in CF. Taking the value of 1.25 
\({\rm \times 10^{-8} erg s^{-1} cm^{-2} sr^{-1}}\)
for the 1 - 2 keV X--ray
background (Hasinger 1992) used by Boyle \etal (1994), the contribution from
AGN rises to between 52\% and 60\% (\(q_{0}=0\)) or 33\% and 38\%
(\(q_{0}=0.5\)).

\begin{figure}
\begin{center}
\leavevmode
\epsfxsize=85mm
\epsfysize=100mm
\epsffile{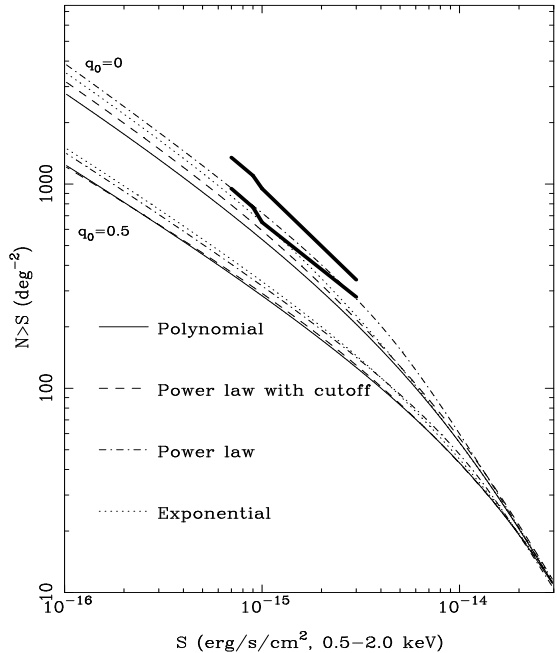}
\caption{Log {\it N} - log {\it S} of the evolution models extrapolated to
faint fluxes. The thick lines represent the \(1 \sigma \) 
confidence limits to the
total log {\it N} - log {\it S} of  all 
sources (i.e. not only AGN) at faint fluxes determined
by fluctuation analysis of a deep {\it ROSAT} field by Barcons \etal (1994).}
\end{center}
\end{figure}

\section{Conclusions}

We have investigated the evolution of the XLF with redshift using a new
sample of 198 X--ray selected AGN with a spectroscopic completeness of 93\% at 
\(3\times 10^{-14} 
{\rm erg\ s^{-1}\ cm^{-2}}\) (0.5 to 2.0 keV). 
We find PLE models, consistent with our data,
in which the XLF declines or ceases to evolve beyond {\it z} \(\sim\) 1.8, 
and we find no
evidence for evolution beyond this redshift from the model independent 
\(< V_{e}/V_{a} >\) test.
We have shown that narrow emission line galaxies at low redshift 
contaminating the AGN sample 
have little effect on the derived evolution properties.
Furthermore, these conclusions are insensitive to the 
 uncertainty in conversion of flux from the 
{\it Einstein} to {\it ROSAT} passbands, although this has a
significant effect on the \(z\) = 0 XLF and evolution rate for \(z\) \(<\) 1.8.

\section{Acknowledgments}

We thank all in the RIXOS team for their work in obtaining and reducing the
data. The RIXOS project has been made possible by the award of International
Time on the La Palma telescopes by the Comit\'e Cient\'ifico Internacional.
This research has made use of data obtained from the UK {\it ROSAT} Data
Archive Centre at the Department of Physics and Astronomy, University of
Leicester (LEDAS) and we would especially like to thank Mike Watson and Steve
Sembay for their kind assistance. We would like to thank the Royal Society for
a grant to purchase equipment essential to the RIXOS project. 
KOM acknowledges the Royal Society for support and 
MJP acknowledges the support of a 
PPARC studentship through the course of this
work. XB and IGS are partially supported by the DGICYT under project
PB92-0501.

\section{References}

\refer Avni Y., and Bahcall J.N., 1980, ApJ, 325, 694

\refer Barcons X., Branduardi-Raymont G., Warwick R.S., Fabian A.C., Mason
K.O., McHardy I.M., Rowan-Robinson M. 1994, MNRAS, 268, 833

\refer Boyle B.J., Shanks T., Peterson B.A.,  
1988, MNRAS, 235, 935

\refer Boyle B.J., Griffiths R.E., Shanks T., Stewart G.C., Georgantopoulos I., 
1993, MNRAS, 260, 49

\refer Boyle B.J., Shanks T., Georgantopoulos I., Stewart G.C.,
Griffiths R.E., 
1994, MNRAS, 271, 639 

\refer Boyle B.J., McMahon R.G., Wilkes B.J., Elvis M., 1995, MNRAS, 272, 462

\refer Branduardi-Raymont G., Mason K.O., Warwick R.S., Carrera F.J.,
Graffagnino V.G., Mittaz J.P.D., Puchnarewicz E.M., Smith P.J., 
Barber C.R., Pounds K.A., Stewart G.C., McHardy I.M., Jones L.R.,
Merrifield M.R., Fabian A.C., McMahon R., Ward M.J., George I.M., 
Jones M.H., Lawrence A., Rowan-Robinson M.,  1994, MNRAS, 270, 947

\refer Chen L.-W., Fabian A.C., Gendreau K.C., submitted to MNRAS

\refer Della Ceca R., Maccacaro T., Gioia I.M., Wolter A., Stocke T.J.,
 1992, ApJ, 389, 491

\refer Dunlop J.S. and Peacock J.A., 1990, MNRAS, 247, 19

\refer Fasano G. and Franceschini A., 1987, MNRAS, 225, 155 

\refer Franceschini A., La Franca F., Cristiani S. and
Martin-Mirones J.M., 1994, MNRAS, 269, 683

\refer Francis P.J., 1993, ApJ, 407, 519

\refer Hasinger G., 1992, in: Barcons X., Fabian A.C. (eds.), The X--ray
Background, Cambridge University Press, p.229

\refer Hasinger G., Burg R., Giacconi R., Hartner G., 
Schmidt M., Trumper J., Zamorani G., 1993, A\&A, 275, 1

\refer Kassiola A., and Mathez G., 1990, A\&A, 230, 255 

\refer Lampton M., Margon B., Bowyer S., 1976, ApJ, 208, 177

\refer Maccacaro T., Gioia I.M., Wolter A., Zamorani G., Stocke J.T.,
1988, ApJ, 326, 680

\refer Maccacaro T., Della Ceca R., Gioia I.M., Morris S.L., 
Stocke J.T., Wolter A., 1991, ApJ, 374, 117

\refer Marshall H.L., Avni Y., Braccesi A., Huchra J., Tananbaum H.,
Zamorani G., Zitelli V., 1984, ApJ, 283, 50

\refer Mason \etal 1995, in preparation

\refer Mittaz \etal 1995, in preparation

\refer Peacock J.A., 1983, MNRAS, 202, 615

\refer Press W.H., Teukolsky S.A., Vetterling W.T., Flannery B.P., 1992,
Numerical Recipes in Fortran, Cambridge University Press, p.640

\refer Primini F.A., Murray S.S., Huchra J., Schild R., Burgh R., Giacconi
R., 1991, ApJ, 374, 440

\refer Stocke J.T., Morris S.L., Gioia I.M., Maccacaro T., Schild R.,
Wolter A., Fleming T.A., Henry J.P., 1991, ApJS, 76, 813
\end{document}